\documentclass[preprint,prd,groupedaddress,showpacs,rotate]{revtex4}

\usepackage{graphicx}
\usepackage{dcolumn}

\usepackage{multirow}

\usepackage[normalem]{ulem}
\usepackage{graphics}
\usepackage{amsmath}
\usepackage{amsthm}
\usepackage{amssymb}
\usepackage{enumerate}
\usepackage{bm,longtable,enumerate}
\usepackage{amsmath,amssymb}

\newcommand{\eqn}{\begin{eqnarray}}
\newcommand{\eqnx}{\end{eqnarray}}

\newcommand\tr{\mathop{\mathrm{Tr}}}

\def\beq{\begin{equation}}
\def\eeq{\end{equation}}
\def\beqa{\begin{eqnarray}}
\def\eeqa{\end{eqnarray}}
\def\ss{\scriptscriptstyle}

\graphicspath{%
    {converted_graphics/}
    {/}
}
\begin{document}


\title{Holographic model for dilepton production in p-p collisions}
\author{C.  A.  Ballon Bayona}
\email{c.a.m.ballonbayona@durham.ac.uk}
\affiliation{Centro Brasileiro de Pesquisas F\'{i}sicas, 
 Rua Dr. Xavier Sigaud 150, RJ 22290-180 -- Brazil \\ and \\
 Centre for Particle Theory, University of Durham, Science Laboratories, South Road, Durham DH1 3LE -- United Kingdom}
\author{Henrique Boschi-Filho }
\email{boschi@if.ufrj.br} 
\affiliation{Instituto de F\'{\i}sica,
Universidade Federal do Rio de Janeiro, Caixa Postal 68528, RJ
21941-972 -- Brazil}
\author{Nelson R. F. Braga}
\email{braga@if.ufrj.br} 
\affiliation{Instituto de F\'{\i}sica,
Universidade Federal do Rio de Janeiro, Caixa Postal 68528, RJ
21941-972 -- Brazil}

\begin{abstract} 
We propose a holographic model for dilepton production in proton-proton collisions
through the exchange of vector mesons. The holographic hard wall model is used to describe the dynamics and interactions of vector mesons and baryons. We estimate the parameters $\lambda,\mu, \nu$ that characterize the angular distribution of the produced dileptons in a region of $q_T^2 << Q^2$, where perturbative QCD presents an effective strong coupling due to large logarithm corrections.
\end{abstract}



\maketitle

\section{ Introduction }

Proton-proton collisions provide a very large amount of information important for the understanding of fundamental interactions. These collisions can produce a large variety of particles through many different processes involving strong and electroweak interactions. Among the observed particles, one can easily distinguish lepton pairs (a lepton and an anti-lepton) usually called dileptons, that can be produced in different ways. Drell and Yan proposed a mechanism, within the parton model, to explain production of dileptons in hadronic collisions \cite{Drell:1970wh,Drell:1970yt}. This mechanism consists on the annihilation of a quark from one hadron with an anti-quark from the other. This annihilation leads to the production a virtual particle (typically a photon) that decays into the lepton pair. 

The analysis of dilepton production is important to understand the internal structure of hadrons.
From the corresponding cross sections one can obtain structure functions related to the
distribution of partons inside the hadron. 

The differential cross section for dilepton production has an angular dependence that can be 
characterized by three scalar parameters: $ \lambda, \mu, \nu $. 
These angular parameters have been studied using perturbative QCD and the parton model originally in
\cite{Kajantie:1978yp,Lam:1978zr,Collins:1978yt,Cleymans:1978ip,Lindfors:1979rc,Lam:1978pu,Collins:1977iv}.
For a more recent discussion see \cite{Boer:2002ju,Boer:2006eq} and references therein. 
These parameters have been measured recently for the case of dimuons by 
FNAL E866/NuSea Collaboration \cite{Zhu:2008sj} using data from 
the collision of 800 GeV beams of protons against a hydrogen target.  
Different hadronic collisions lead to different angular parameters. 
See for instance the case of proton-deuteron  in \cite{Zhu:2006gx}. 

The perturbative QCD calculations for the Drell Yan cross section work well when the transverse momentum of the dilepton is large with respect to the dilepton mass. However, when the transverse momentum is small, 
the perturbative series involves the product of the (small) coupling constant $\alpha_S$  with large logarithm corrections. This combination leads to a series with a strong effective coupling. 
So, the standard perturbative approach is not reliable in this regime. In this case one needs a resummation to all orders, as discussed in \cite{Boer:2006eq}. 
 
Recently, alternative approaches to gauge theories at strong coupling were developed  based on gauge string dualities inspired in the AdS/CFT correspondence \cite{Maldacena:1997re,Gubser:1998bc,Witten:1998qj}.
In particular these dualities lead to some holographic models to describe non perturbative aspects of 
QCD known as AdS/QCD (see for instance 
\cite{Polchinski:2001tt,Polchinski:2002jw,BoschiFilho:2002vd,
BoschiFilho:2002ta,deTeramond:2005su,Erlich:2005qh,
Hong:2006ta,Grigoryan:2007vg,Sakai:2004cn,Sakai:2005yt}). 

In this article we propose a holographic model to calculate contributions to dilepton production 
through the decay of a virtual photon in proton-proton collisions. 
Inspired by vector meson dominance, in our model the virtual photon giving rise to the dilepton comes from the decay of a vector meson. 
This vector meson is produced by the annihilation of two other vector mesons emitted by the protons.  
The dynamics and interactions of baryons and vector mesons are described using the AdS/QCD hard wall model \cite{Polchinski:2001tt,Polchinski:2002jw,BoschiFilho:2002vd,BoschiFilho:2002ta}. 
Hadrons correspond to modes of a Kaluza-Klein expansion of five dimensional fields 
living in an AdS slice. The size of the slice represents a mass gap in the 4-d effective theory. 
The hadronic masses are determined by the five dimensional wave functions and the boundary conditions 
while the effective coupling constants arise from the integration of interaction terms in the 5-d action.
These masses and couplings are used to calculate the scattering amplitude relevant for the process of dilepton production.  For simplicity, we consider only final hadronic states with spin 1/2. Using this model we find the angular distribution parameters $ \lambda, \mu, \nu $, for kinematical regimes where the dilepton transverse momenta are small. 

In section {\bf 2} we review dilepton production in proton-proton collisions. In section {\bf 3} we 
calculate, within the hard wall model, the fermion and vector meson masses and couplings relevant for 
our model. We present in section {\bf 4} our model for inclusive dilepton production and estimate 
the parameters $\lambda, \mu, \nu$, that characterize the angular dependence of the differential 
cross section, for  kinematical regimes compatible with those analysed by FNAL E866/NuSea Collaboration.

\section{Inclusive dilepton production in P-P collisions}

The production of dilepton from a proton-proton collision through the decay of a virtual photon is represented in Figure 1. Two protons with initial momenta $P_1$ and $P_2$ interact producing a time-like virtual photon with momentum $q$ plus some additional hadronic states which are not observed. The virtual photon decays into a lepton and anti-lepton with momenta $k_1$ and $k_2$. This section is  based on the extensive discussion of dilepton production presented in ref. \cite{Lam:1978pu}. For a more recent study of angular distribution in Drell Yan process, see also 
\cite{Boer:2002ju,Boer:2006eq}. 

The differential cross section, for the unpolarized case, can be written as 
\beq
d \sigma =   \frac{e^4}{ (q^2)^2 s } \, W^{\mu \nu} L_{\mu \nu} \, 
\frac{d^3 \vec{k}_1}{(2 \pi)^3 2 |\vec{k}_1| }\frac{d^3 \vec{k}_2}{(2 \pi)^3 2 |\vec{k}_2| }  \,, 
\eeq
where $s=-(P_1 + P_2)^2$ and $q^2=-m_{\gamma}^2$ with $m_\gamma > 0$ the virtual photon mass. The lepton masses were neglected with respect to their momenta and the proton masses were neglected with respect to the center of mass energy $\sqrt{s}$. 

The hadronic tensor $W^{\mu \nu}$ is expressed in terms of the matrix elements of the electromagnetic hadronic current as \footnote{Note that the  tensor $W^{\mu \nu}$  of ref.  \cite{Lam:1978pu} corresponds to our  $W^{\mu \nu}$ multiplied by $s$ and replacing $q$ by $-q$, while the tensor  $L_{\mu \nu}$ of ref.  \cite{Lam:1978pu} is our tensor $L_{\mu \nu}$ multiplied by $2q^2$} 
\beq
W^{\mu \nu} = \frac14 \sum_{S_{H1}} \,  \sum_{S_{H2}} \int d^4 x e^{-i q \cdot x} 
\langle P_1 , S_{H1} , P_2, S_{H2} \vert \left [ J^\mu_H (x), J^\nu_H (x) \right ] 
\vert P_1 , S_{H1} , P_2, S_{H2}  \rangle \,,
\eeq
where $S_{H1}$ and $S_{H2}$ are the spins of the initial hadrons. On the other hand, the leptonic tensor $L_{\mu \nu}$ is defined in terms of the leptonic current as
\beqa
L_{\mu \nu} &=& \sum_{S_{L1}} \sum_{S_{L2}} \langle k_1 , S_{L1} \vert J_\mu^{L}(0) \vert-k_2 , S_{L2} \rangle  \langle - k_2 , S_{L2} \vert J_\nu^{L}(0) \vert k_1 , S_{L1} \rangle \cr 
&=& 4 \left [ k_1 \cdot k_2 \, \eta_{\mu \nu} - k^1_\mu k^2_\nu - k^1_\nu k^2_\mu \right ] \,, 
\eeqa
where $S_{L1}$ and $S_{L2}$ are the spins of the leptons and $\eta_{\mu \nu} = {\rm diag}(-,+,+,+)$.

The most general tensor $W^{\mu \nu}$ that satisfies hermiticity, parity constraints and gauge invariance can be decomposed as 
\beqa
\! \! W^{\mu \nu} \! \! &=& \left (\eta^{\mu \nu} - \frac{q^\mu q^\nu}{q^2} \right) W_1   \cr 
 &+&  \frac{1}{s} \!\left [ P_1^\mu - \frac{P_1 \cdot q}{q^2}q^\mu + P_2^\mu - \frac{P_2 \cdot q}{q^2}q^\mu \right ]\! \left [ P_1^\nu - \frac{P_1 \cdot q}{q^2}q^\nu + P_2^\nu - \frac{P_2 \cdot q}{q^2}q^\nu \right ] \!W_2 \cr 
&-& \frac{1}{s} \! \left \{ \! \left[P_1^\mu - \frac{P_1 \cdot q}{q^2}q^\mu \right] \! \left[P_1^\nu - \frac{P_1 \cdot q}{q^2}q^\nu \right] -  \left[P_2^\mu - \frac{P_2 \cdot q}{q^2}q^\mu \right]\! \left[P_2^\nu - \frac{P_2 \cdot q}{q^2}q^\nu \right] \!  \right \} \! W_3 \cr 
&+& \! \frac{1}{s} \!  \left [ P_1^\mu - \frac{P_1 \cdot q}{q^2}q^\mu - P_2^\mu + \frac{P_2 \cdot q}{q^2}q^\mu \right ] \! \left [ P_1^\nu - \frac{P_1 \cdot q}{q^2}q^\nu - P_2^\nu + \frac{P_2 \cdot q}{q^2}q^\nu \right ] \! W_4 \, , \label{Wstructure}
\eeqa

\begin{figure}\begin{center}
\setlength{\unitlength}{0.06in}
\vskip 3.cm
\begin{picture}(0,0)(15,0)
\rm
\thicklines
\put(-1,14.5){$P_1$}
\put(3.5,15){\line(3,-2){10.4}}
\put(3.5,15){\vector(3,-2){6}}
\put(-1,-6){$P_2$}
\put(4.1,-3.1){\line(3,2){10.3}}
\put(4.1,-3.1){\vector(3,2){6}}
\put(16,6){\circle{5.5}}
\put(25,15){$X$}
\put(17.2,8.5){\line(3,4){5}}
\put(17.2,8.5){\vector(3,4){2.6}}
\put(17.7,8.2){\line(1,1){6}}
\put(17.7,8.2){\vector(1,1){3}}
\put(18.2,7.8){\line(3,2){7.2}}
\put(18.2,7.8){\vector(3,2){3.3}}
\put(25,-6){$Y$}
\put(17.2,3.4){\line(3,-4){5}}
\put(17.2,3.4){\vector(3,-4){2.5}}
\put(17.7,3.8){\line(1,-1){6}}
\put(17.7,3.8){\vector(1,-1){3}}
\put(18.2,4.2){\line(3,-2){7.5}}
\put(18.2,4.2){\vector(3,-2){3.7}}
\put(23.5,3){$\gamma$}
\multiput(3.7,-2)(4.5,0){2}{
\bezier{300}(15,7.5)(16.5,9.7)(17.5,7.5)
\bezier{300}(17.5,7.5)(18.5,6.5)(19.5,7.5)}
\put(33,11.5){$k_1$}
\put(27.7,5.5){\line(1,1){5}}
\put(27.7,5.5){\vector(1,1){2.8}}
\put(33,-1.5){$k_2$}
\put(27.7,5.5){\line(1,-1){5}}
\put(27.7,5.5){\vector(1,-1){2.8}}
\end{picture}
\vskip 1.cm
\parbox{4.1 in}
{\caption{Illustrative diagram for a Drell-Yan scattering.}}
\end{center}
\end{figure}
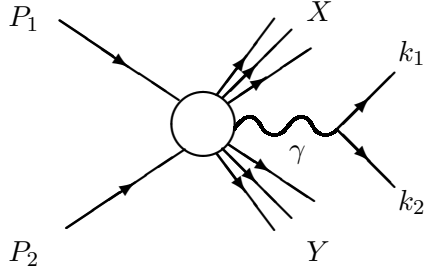
\vskip .5cm

\noindent where $W_1, W_2, W_3, W_4$ are the invariant hadronic structure functions that contain the relevant information for the dilepton cross section. These structure functions depend on four scalar variables that are combinations
of the momenta $P_1, P_2 $ and $q$. 

It is useful to introduce the helicity structure functions 
\beq
W_T = W_{1,1} \, \,  , \, \,  W_L = W_{0,0} \, \, , \, \,  W_\Delta = \frac{1}{\sqrt{2}}(W_{0,1}+W_{1,0}) \, \, , \, \, 
W_{\Delta \Delta} = W_{1,-1} \,,
\eeq
where 
\beq
W_{\sigma,\sigma'} = \eta^\mu_{_{(\sigma)}} \eta^{\ast \nu}_{_{(\sigma')}} W_{\mu \nu} \qquad  
\qquad \sigma,\sigma'=(-1,0,1)\,, 
\eeq
and $\eta^\mu_{_{(\sigma)}}(q)$ are the polarization vectors of the virtual photon, in its rest frame, defined in terms of Cartesian unit spatial vectors $X^\mu=(0, \vec{X})$, $Y^\mu= (0, \vec{Y})$, $Z^\mu= (0, \vec{Z})$ as  
\beq
\eta^\mu_{_{(0)}} = Z^\mu \quad ; \quad  \eta^\mu_{_{(\pm 1)}} = \frac{1}{\sqrt{2}}( \mp X - i Y)^\mu \,.
\eeq

 The hadronic tensor can also be decomposed in terms of the helicity structure functions as 
\beqa
W^{\mu \nu} &=&  \left (\eta^{\mu \nu} - \frac{q^\mu q^\nu}{q^2} \right) (W_T + W_{\Delta \Delta}) - 2 X^\mu X^\nu W_{\Delta \Delta} \cr 
&+& Z^\mu Z^\nu (W_L - W_T - W_{\Delta \Delta} ) - (X^\mu Z^\nu + X^\nu Z^\mu ) W_\Delta \,, \label{Whelicity}
\eeqa
so that the differential cross takes the form (in the photon rest frame)
\beqa
d \sigma &=& \frac{e^4}{8 (2\pi)^6 \, m_\gamma^2 s} \Big [ W_T ( 1 + \cos^2\theta) + W_L (1 - \cos^2 \theta) + W_\Delta \sin 2 \theta \cos \phi \cr 
&&  \qquad\qquad\qquad  +\, W_{\Delta \Delta} \sin^2\theta \cos 2\phi  \Big ] \sin \theta d \theta d \phi \, d^4 q \, , 
\eeqa
where $\theta$ and $\phi$ are the spherical angular coordinates for the vector $\vec{k}_1$ with respect to the Cartesian system $X,Y,Z$ : $\vec{k}_1 = |\vec{k}_1| (\sin \theta \cos \phi , \sin \theta \sin \phi, \cos \theta)$. 

In order to compare with experimental results it is interesting to introduce  parameters that characterize the angular dependence of the cross section. These parameters are defined by 
\beq
\lambda = \frac{W_T-W_L}{W_T+W_L} \quad , \quad \mu = \frac{W_\Delta}{W_T+W_L} \quad , 
\quad \nu = \frac{2 W_{\Delta \Delta}}{W_T+W_L} \,,  \label{lambdamunu}
\eeq
so that 
\beq
d \sigma \sim 1 + \lambda \cos^2 \theta + \mu \sin 2 \theta \cos \phi + \frac{\nu}{2}\sin^2\theta \cos 2\phi \,.
\eeq

Note that the helicity structure functions depend on the choice of the unit vectors $X,Y,Z$ which must be  defined  
in terms of the momenta $P_1$, $P_2$ and $q$. There are different possible choices as discussed in  \cite{Lam:1978pu}. 
Here we will follow the Collins-Soper frame \cite{Collins:1977iv},
 which is 
\beqa
Z^\mu &=& \frac{1}{\sqrt{s} (m_\gamma^2 + q_T^2)^{1/2
}} \Big \{ q_p \left [ P_1^\mu - \frac{P_1 \cdot q}{q^2}q^\mu + P_2^\mu - \frac{P_2 \cdot q}{q^2}q^\mu \right ] \cr 
&+& q_P \left [ P_1^\mu - \frac{P_1 \cdot q}{q^2}q^\mu - P_2^\mu + \frac{P_2 \cdot q}{q^2}q^\mu \right ] \Big \} \,, \cr \cr 
X^\mu &=& - \frac{m_\gamma }{\sqrt{s} q_T (m_\gamma^2 + q_T^2)^{1/2
}} \Big \{ q_P \left [ P_1^\mu - \frac{P_1 \cdot q}{q^2}q^\mu + P_2^\mu - \frac{P_2 \cdot q}{q^2}q^\mu \right ] \cr 
&+& q_p \left [ P_1^\mu - \frac{P_1 \cdot q}{q^2}q^\mu - P_2^\mu + \frac{P_2 \cdot q}{q^2}q^\mu \right ] \Big \} \,, \cr \cr 
Y^\mu &=& (0 , \vec{Y}) = (0, \vec{Z} \times \vec{X})\,,  \label{unitvectors}
\eeqa
where 
\beq
q_p = \frac{ q \cdot (P_1 - P_2) }{\sqrt{s}} \quad , \quad q_P = -\frac{ q \cdot (P_1 + P_2)}{\sqrt{s}} \quad , 
\quad q_T = ( q_P^2 - q_p^2 - m_\gamma^2 )^{1/2} \,.
\label{qes}
\eeq

Using the coordinate choice given by (\ref{unitvectors}) in eq. (\ref{Whelicity}) and comparing with the expansion 
of the hadronic tensor in terms of the invariant structure functions given in eq. (\ref{Wstructure}) one finds 
\beqa
W_T &=& W_1 + \frac{q_T^2}{2 m_\gamma^2 } \, \frac{q_P^2 W_2 + q_P q_p W_3 + q_p^2 W_4 }{m_\gamma^2 + q_T^2} \cr 
W_L &=& W_1 + \frac{q_p^2 W_2 + q_P q_p W_3 + q_P^2 W_4 }{m_\gamma^2 + q_T^2} \cr 
W_\Delta &=& -\frac{q_T}{ m_\gamma } \, \frac{q_P q_p (W_2 + W_4) +(1/2)(q_P^2 + q_p^2) W_3}{m_\gamma^2 + q_T^2} \cr 
W_{\Delta \Delta} &=& - \frac{q_T^2}{2 m_\gamma^2 } \, \frac{q_P^2 W_2 + q_P q_p W_3 + q_p^2 W_4 }{m_\gamma^2 + q_T^2} \,. 
\label{LamTungeq}
\eeqa

Using perturbative QCD and the parton model, one can calculate the angular distribution parameters $\lambda$ and $\nu$ for large $q_T$, as functions of $q$ and $q_T$, finding in the Collins-Soper frame (in our signature where $q^2 < 0 $) \cite{Collins:1978yt}
\begin{equation}
\lambda_{pert} = \, \frac{q^2 +\frac{1}{2} q_T^2 }{q^2 - \frac{3}{2} q_T^2 }\,,\quad \qquad
\nu_{pert} = \,\frac{ q_T^2 }{ - q^2 + \frac{3}{2} q_T^2 }\,. \label{perturb_results}
\end{equation}

\noindent 
The angular parameter $\mu_{pert}$ can not be written simply as a function of $q$ and $q_T$, even for large $q_T$. Rather, it involves integrals over parton distribution functions, which are not {a priori} known \cite{Boer:2006eq}.

In the following sections we develop a holographic model for calculating the invariant structure functions $W_1,W_2,W_3,W_4$ and then we estimate the parameters $\lambda , \mu,\nu$ for kinematical regimes with small $q_T$.

The kinematical regimes that we will investigate are in the region analyzed recently in ref. \cite{Zhu:2008sj},  considering dimuons produced in collisions of 800 GeV beams of protons against a hydrogen target. In this reference de range of dimuon masses analysed was  $ 4.5 < m_{\mu\mu} < 15 {\rm GeV} $ and the 
transverse momenta is in the region $0 < q_T < 4 \, {\rm GeV}$  and $ 0 < x_F < 0.8 $. The mean values found for the angular distribution parameters were:
\beq 
\langle \, \lambda \,\rangle = 0.85 \quad;\quad  \langle \, \mu \, \rangle = -0.026 \quad ;\quad  
\langle \,\nu  \, \rangle = 0.04 \,\,.
\eeq

In order to compare our results with those of ref. \cite{Zhu:2008sj} we identify 
$  m_\gamma = m_{\mu\mu} $.


\section{Vector mesons and baryons in the hard wall model}

In this paper we consider the production of dileptons in proton-proton collisions
through the exchange of vector mesons. So we need first to describe vector mesons and baryons and their interactions in the hard wall model. For simplicity, we consider that each final hadronic state is just a spin 1/2 baryon. For previous discussions of hadrons in AdS/QCD models see, for instance, 
 \cite{deTeramond:2005su,Erlich:2005qh,Hong:2006ta,Grigoryan:2007vg}. 

The hard wall model consists on a 5-d theory living in an AdS$_5$ slice with metric
\begin{equation}
ds^2=\frac{R^2}{z^2}[-dt^2+d{\vec{x}}^2+dz^2]\,,
\label{AdSmetric}
\end{equation}
where $0<z<z_0 = 1/\Lambda$ and $\Lambda$ is an IR energy scale for the dual 4-d effective theory. 
The physical spectrum of the hadronic particles is obtained after imposing boundary conditions at $z=z_0$.

\subsection{Wave functions of vector mesons}

Consider the action for the non-Abelian 5-d gauge fields in the presence of a gauge-fixing term 
\begin{equation}
S= \kappa \int d^4xdz \sqrt{-g} \tr \left \{ -\frac14F_{MN}F^{MN} - \frac{1}{2 \xi} \left ( \frac{1}{\sqrt{-g}} \partial_M (\sqrt{-g} A^M ) \right )^2 \right \} \,,
\end{equation}
where $F_{MN}=\partial_M A_N - \partial_N A_M + [A_M,A_N]$, with $A_M=A_M^aT^a$,
and $T^a$ are the generators of the $SU(N_f)$ flavour group. 
The constant $\kappa$ will be related in the following to the normalization of the gauge fields $A^M$. As we are going to show at the end of section IV, this constant  will not contribute to the angular parameters $\lambda$, $\mu$, and $\nu$.

For the AdS metric (\ref{AdSmetric}) and setting $A_z=0$ we find
\begin{equation}
S= \kappa \int d^4xdz \sqrt{-g}\,h^2\,
\tr \left [- \frac14 \eta^{\mu\sigma}\eta^{\nu\rho}F_{\mu\nu}F_{\sigma\rho}
- \frac12 \eta^{\mu\nu} \partial_z A_{\mu}\partial_z A_\nu - \frac{1}{2 \xi} ( \eta^{\mu \nu} \partial_\mu A_\nu )^2 \right ] \,, 
\end{equation}

\noindent 
where $h=h(z)=z^2/R^2$ and $\eta_{\mu\nu}$ is the 4-d Minkowski metric.

Since the $z$-coordinate is compact, we consider a Kaluza-Klein expansion for the gauge field
\begin{equation}
A_\mu(z,x) \,=\, f^{0}(z)\, a_\mu (x) + \sum_{n=1}^{\infty}f^{n}(z)\, v_\mu^{n}(x) \,,
\end{equation}
where the modes $f^0, f^n$ satisfy Neumann boundary conditions at the IR cut off $z=z_0$ 
and regularity conditions $f^0(z)=1$, $f^n(z)=0$ at the spacetime boundary $z=0$.

It is important to remark that the infrared Neumann boundary condition arises from the gauge invariant boundary condition $F_{z \mu} = 0$. 
The latter is a necessary condition to preserve the $U(N_f)$ gauge symmetry of the  vector field in the bulk. An infrared
 Dirichlet condition is not allowed because it breaks the bulk gauge invariance. 
For a discussion see ref.\cite{Hirn:2005nr}.

Imposing the conditions
\beqa
\kappa  \int_0^{z_0} dz \sqrt{-g} \, h^2 f^{n}(z) f^{m}(z) &=& \delta^{n m}  \quad , \label{norm-fn}
 \\ \cr 
\frac{1}{\sqrt{-g} \, h^2} \, \partial_z \left[ \sqrt{-g} \, h^2 \partial_z f^{n}(z) \right ] 
&=& - m_n^2  \, f^n (z) \quad , \label{eqmovfn}
\eeqa
for the modes $n={1,2,3,\dots}$, and the condition 
\beq
 \partial_z \left[ \sqrt{-g} \, h^2 \partial_z f^{0}(z) \right ] = 0 \label{eqmovf0}
\eeq
for the non-normalizable zero mode, we obtain a 4-d effective action
\beqa
S &=&  \int d^4 x \tr \Big\{\sum_{n=1}^{\infty} \Big[ -\frac14 \eta^{\mu \sigma} \eta^{\nu \rho} v^n_{\mu \nu} v^n_{\sigma \rho}
 -\frac{1}{2\xi} (\eta^{\mu\nu} \partial_\mu v_\nu^n )^2   -\frac{ m_n^2}{2}  \eta^{\mu \nu}  v_\mu^n v_\nu^n  \cr
&& - \frac{d_{n}}{2}   \eta^{\mu \sigma} \eta^{\nu \rho}  a_{\mu \nu} v^n_{\sigma \rho}   
 - \frac{d_n}{\xi} \eta^{\mu \nu} \eta^{\sigma \rho} \partial_\mu a_\nu \partial_\sigma v_\rho^n \Big] \cr
 && -\frac{d_0}{4}   \, \eta^{\mu \sigma} \eta^{\nu \rho} a_{\mu \nu}a_{\sigma \rho}  - \frac{d_0}{2\xi} (\eta^{\mu \nu} \partial_\mu a_\nu)^2  \Big\}
+ S_{int.} \,, 
\eeqa
where $S_{int.}$ represents the interactions terms, we defined 
\beqa
d_0 \equiv  \kappa \int dz \sqrt{-g} \, h^2 (f^0)^2 \qquad ; \qquad 
d_{n} \equiv \kappa \int dz \sqrt{-g} \, h^2 f^0 f^n \, 
\eeqa
and $a_{\mu \nu}\equiv \partial_\mu a_\nu - \partial_\nu a_\mu$,   
$\,v^n_{\mu \nu}\equiv \partial_\mu v^n_\nu - \partial_\nu v^n_\mu$ are the kinetic parts of the 4-d gauge field strengths.

In order to diagonalize the kinetic terms we redefine the vector field
\beq
 v^n_\mu (x) = \tilde v^n_\mu (x) - d_n a_\mu (x) \, ,
\eeq
so that the action takes the form 
\beqa
S &=& \int d^4 x \sum_{n=1}^{\infty} \tr \Big \{  
\frac12 \tilde v_\mu^n \left [  \eta^{\mu \nu} (\partial^2 -m_n^2) 
+ \left(\frac{1}{\xi} - 1 \right)\eta^{\mu \sigma} \eta^{\nu \rho} \partial_\sigma \partial_\rho \right ]\tilde v_\nu^n +  g_{v^n} \eta^{\mu \nu} \tilde v_\mu^n a_\nu \,  \Big \} 
 \cr
&&  
+ S_{int.}  \,+ \, \dots \,, \label{4d-action}
\eeqa
with $g_{v^n}=m_n^2 d_n$ and the dots represent the divergent terms arising from $a_\mu$.
The fields $\tilde v^n_{\mu}$ are interpreted as vector mesons with mass $m_n$ while the 
field $a_\mu$ is interpreted as the photon which decompose into vector mesons with decay constant  $g_{v^n}$. 
This way the hard wall model realizes vector meson dominance. 

Note that the operator between brackets in the vector meson kinetic term  depends on the parameter $\xi$. 
The 4-d vector meson propagator is given by the inverse of this operator in momentum space and takes the form 
\beq
\Delta^{\mu \nu}( P, m_n^2) = \frac{1}{P^2 + m_n^2 } \left [ \eta^{\mu \nu}   - (1-\xi) \frac{P^\mu P^\nu }{P^2 + \xi m_n^2 } \right ] \,. 
\label{vmpropagator}
\eeq

In order to calculate the masses and couplings appearing in eq. (\ref{4d-action}), we now consider the solutions of eq. (\ref{eqmovfn}) with the chosen boundary conditions
\begin{equation}
f^n(z)=c_n z J_1(m_n z) \,,
\label{fn}
\end{equation}
where $m_n z_0$ are zeros of the Bessel function $J_0(w)$, 
implied by the Neumann boundary condition over $f^n(z)$ at $z=z_0$. 
The normalization condition (\ref{norm-fn})   implies that
\begin{equation}
c_n=\sqrt{\frac{2}{\kappa R}}\, \frac 1{z_0 |J_1(m_n z_0)|} \,. 
\label{cn}
\end{equation}

The solution for the zero mode eq. (\ref{eqmovf0})  satisfying the corresponding boundary conditions 
is $f^0(z)=1$.

The coupling constant $g_{v^n}$ takes the form
\begin{eqnarray}
g_{v^n} &=& m_n^2 \kappa \int_0^{z_0} dz \sqrt{-g}\, h^2\, f^n(z)\cr
&=& \kappa R \lim_{\epsilon \to 0} \frac 1{\epsilon} 
\left. \partial_z f^n(z)\right|_{{z=\epsilon}} \;
=\;  \kappa R \, c_n \, m_n \,. \label{gvn}
\end{eqnarray}

Now we consider the interaction Lagrangian for three vector mesons 
\begin{eqnarray}
S_{VVV} &=& -{\kappa} \int d^4xdz \sqrt{-g}\,h^2\, \eta^{\mu\alpha}\eta^{\nu\beta} 
\tr \left\{[A_\mu,A_\nu] \, \partial_\alpha A_{\beta} \right\} \cr 
&=& - \, \sum_{n,m,\ell} g_{v^nv^mv^{\ell}} \int d^4x \, \eta^{\mu\alpha}\eta^{\nu\beta} 
\tr \left\{[\tilde v^n_\mu, \tilde v^m_\nu] \, \partial_\alpha \tilde v^{\ell}_{\beta} \right\}  + \, \dots \,, 
\end{eqnarray} 
where the triple coupling is 
\begin{equation}
 g_{v^nv^mv^{\ell}} = \kappa \int dz \sqrt{-g} \, h(z)^2 f^n(z) f^m(z) f^{\ell}(z) \,, \label{gvvv}
\end{equation}
and the dot terms are divergent terms arising from the photon $a_\mu$. 

In order to calculate the masses and couplings for the vector mesons, it is important to discuss their dependence on the parameters of the model. 
The masses depend on the IR scale $\Lambda = 1/z_0$ (the same will happen in the fermionic case). 
We will fix this scale $\Lambda $ in section IV using the mass of the $\rho$ meson.  On the other hand, the couplings
carry a dependence on the product $\kappa R$. In particular, using eqs. (\ref{cn}) and (\ref{gvn}), one finds that the couplings $g_{v^n} $ are proportional to $\sqrt{ \kappa R  } \Lambda^2 $,  while from 
eqs.  (\ref{fn}) , (\ref{cn}) and (\ref{gvvv}) one concludes that the triple couplings $ g_{v^nv^mv^{\ell}} $ contain a factor $1/\sqrt{ \kappa R  } $.  The factor $\kappa R$ will appear as a multiplicative factor in the hadronic tensor. So, as we will discuss in section IV,  it will not contribute to the angular parameters, that involve only ratios of structure functions. So, we do not need to fix a value for this quantity. 
We show in the appendix A some numerical results for the masses and coupling of the 
vector mesons, up to these factors.

It is important to remark that the approach to vector mesons 
in the hard wall model that we considered here is very similar to the one presented in 
\cite{Sakai:2004cn,Sakai:2005yt} within the D4-D8 brane model.


\subsection{Wave functions of spin 1/2 baryons}

In order to describe  spin 1/2 states within the hard wall model, 
we start with the 5-d Dirac action
\begin{eqnarray}
S_F = {\kappa_F} \int d^4xdz \sqrt{-g}\,\bar\psi (D-\bar M)\psi  \label{Diracaction}\,, 
\end{eqnarray} 
where $\psi$ is a 5-d spinor with mass $\bar M$  and the covariant derivate is defined as
\begin{equation}
 D = \frac{z}{R} \left [ \gamma^5 \partial_z +  i \, \eta^{\mu\nu}  \gamma_\mu \partial_\nu \right ] 
- \frac{2}{R} \gamma^5 \,, 
\end{equation}
with $\gamma_\mu$ the 4-d Dirac gamma matrices and $\gamma^5$ the chirality matrix. 
 The constant $\kappa_F$ will be related in the following to the normalization of the fermionic fields $\psi$ and $\bar\psi$. As we are going to show at the end of this  section, this constant will not contribute to the angular parameters $\lambda$, $\mu$,  $\nu$ or to the hadronic tensor $W^{\mu\nu}$.

We consider the Kaluza-Klein expansion 
\beqa
\psi &=& \sum_{n=1}^{\infty} \left [ \phi^n(z) {\cal P}_+ + \tilde \phi^n (z) {\cal P}_- \right ] u^n (x) \cr 
\bar \psi &=&  \sum_{n=1}^{\infty} \bar u^n(x) \left [ \phi^n(z) {\cal P}_-  + \tilde \phi^n (z) {\cal P}_+ \right ] \, ,
\eeqa
with ${\cal P}_{\pm}=(1/2)(1 \pm \gamma^5)$. Imposing the conditions 
\beqa
\kappa_F \int_0^{z_0} \sqrt{-g} \frac{z}{R} \phi^n(z) \phi^m(z) &=& \kappa_F \int_0^{z_0} \sqrt{-g} \frac{z}{R} \tilde \phi^n(z) \tilde \phi^m(z) = \delta_{nm} \, , \label{normfermion} \\
 \cr 
\left [ z \partial_z  - (2 + \bar M R) \right ] \phi^n &=& - z M_n \tilde \phi^n \, ,\cr \cr
\left [ -z \partial_z  + (2 - \bar M R)  \right ] \tilde \phi^n &=& - z M_n \phi^n \,, \label{eqmovfermion}
\eeqa
we find the 4-d effective action 
\beq
S_F = \sum_{n=1}^{\infty} \int d^4 x  \bar u^n (x) \left [ i \, \eta^{\mu \nu} \gamma_\mu \partial_\nu  - M_n   \right ] u^n(x) \,, 
\eeq
where $M_n$ are the masses of the baryonic states of spin $1/2$ in the 4-d theory. $M_1$ is identified with the proton mass and $M_n$ 
with $n=2,3,\dots$, correspond to excited states.  

The normalizable solutions to the eqs. (\ref{eqmovfermion}) are 
\beqa
\phi^n(z) &=& N_n z^{5/2} J_{\bar M R  - 1/2} (M_n z) \, , \cr 
\tilde \phi^n(z) &=& \tilde N_n  z^{5/2} J_{\bar M R + 1/2 } (M_n z) \,.
\label{solufermions}
\eeqa
According to the AdS/CFT correspondence, the dimension of the boundary fermionic operator is related to the mass of the bulk 
fermionic field by 
\beqa
\Delta = \bar M R + 2 \, .
\eeqa
Baryon states in QCD are associated to a fermionic operator with dimension $\Delta = 9/2$. For this reason we fix the 5-d mass to $\bar M R = 5/2$. 

We have two possible boundary conditions in the wall $z = z_0$. We can set $\phi^n(z_0) = 0$ so that the baryon masses are given by  $M_n z_0 = \chi_n$ where $\chi_n$ are the zeros of the Bessel function $J_{\bar M R  - 1/2}$. From the normalization condition we obtain 
\beqa
N_n = \tilde N_n = \sqrt{\frac{2}{\kappa_F R^4 }} \frac{1}{z_0 | J_{\bar M R  + 1/2} (M_n z_0)   |} \,.
\label{normafermions}
\eeqa
Alternatively, we can choose $\tilde \phi^n (z_0)= 0$ so that the masses are given by  $M_n z_0 = \bar \chi_n$ where $\bar \chi_n$  are the zeros of $J_{\bar M R  + 1/2}$. The normalization constants in that case take the form
\beqa
N_n = \tilde N_n = \sqrt{\frac{2}{\kappa_F R^4 }} \frac{1}{z_0 | J_{\bar M R  - 1/2} (M_n z_0)   |} \,.
\eeqa

In this work we choose the boundary condition $\phi^n(z_0) = 0$ to calculate the structure functions and the angular parameters
 $\lambda, \mu, \nu $ for several kinematical regimes. We also estimate the error associated with this choice of boundary condition  by calculating $\lambda, \mu, \nu $ for some particular kinematical regimes with the alternative condition $\tilde \phi^n (z_0)= 0$.


\subsection{Interaction of baryons and vector mesons}

We can describe the interaction of two fermions and one vector meson considering the 5-d action 
\beq
S_{FFV} = \kappa_F \int d^4 x d z \sqrt{-g}  \frac{z}{R} \eta^{\mu \nu} \bar \psi \gamma_\mu A_\nu \psi  
\eeq
that comes from imposing invariance of the action (\ref{Diracaction}) with respect to 4-d gauge transformations. Using the Kaluza-Klein expansions for the fields we find
\beqa
S_{FFV} = \sum_{n,m,\ell} \int d^4 x \, \eta^{\mu \nu} \, \bar u^n  \gamma_\mu \left[ g^+_{\bar u^n  v^m u^\ell}  
{\cal P}_+  + g^-_{\bar u^n v^m u^\ell}  {\cal P}_-  \right]  \tilde v^m_\nu  u^\ell \,+\, \dots \,, 
\eeqa
where 
\beqa
g^+_{\bar u^n v^m u^\ell} = \kappa_F \int_0^{z_0} dz \sqrt{-g} \frac{z}{R}  
\phi^n(z) f^m(z) \phi^\ell(z) \,, \cr 
g^-_{\bar u^n v^m u^\ell} = \kappa_F \int_0^{z_0} dz \sqrt{-g} \frac{z}{R}  
\tilde \phi^n(z) f^m(z) \tilde \phi^\ell(z) \,. \label{ghhv}
\eeqa

From eqs. (\ref{solufermions}) and  (\ref{normafermions}) one finds that the above couplings are actually independent of the parameter $\kappa_F $. Thus, $\kappa_F $ does not contribute to any of our results and does not need to be fixed. On the other hand, the fermionic couplings (\ref{ghhv}) are proportional to $1/\sqrt{ \kappa R  } $. However, as we already mentioned in the case of vector mesons, we do not need to fix this quantity either, since it will not contribute to the angular parameters $\lambda,\mu, \nu$.  

We show in appendix A  some of the numerical results for the masses and couplings of vector mesons and baryons that we used in our calculations.

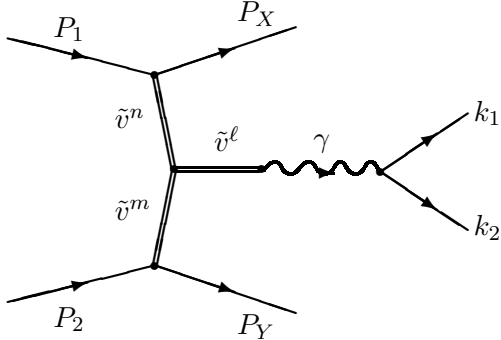
\begin{figure}
\begin{center}
\vskip 3.cm
\begin{picture}(0,0)(25,0)
\setlength{\unitlength}{0.04in}
\rm
\thicklines 
\put(-3.8,20){\circle*{1}}
\put(-17,25){$P_1$}
\put(-3.8,20){\line(-4,1){18.5}}
\put(-23,24.8){\vector(4,-1){10.5}}
\put(7,27){$P_X$}
\put(-3.8,20){\line(3,1){18.5}} 
\put(-3.8,20){\vector(3,1){10.5}}
\put(-3.8,-5){\circle*{1}}
\put(-17,-13){$P_2$}
\put(-3.8,-5){\line(-4,-1){18.5}}
\put(-23,-9.8){\vector(4,1){10.5}}
\put(7,-14){$P_Y$}
\put(-3.8,-5){\line(3,-1){18.5}} 
\put(-3.8,-5){\vector(3,-1){10.5}}
\put(-9,13){$\tilde v^n$}
\put(-1.0,7.4){\line(-1,5){2.5}}
\put(-1.5,7.4){\line(-1,5){2.5}} 
\put(-9,1){$\tilde v^m$}
\put(-1.0,7.4){\line(-1,-5){2.5}}
\put(-1.5,7.4){\line(-1,-5){2.5}}
\put(10.2,7.6){\circle*{1}}
\put(-1.2,7.8){\line(1,0){11.1}}
\put(-1.2,7.3){\line(1,0){11.1}}
\put(-1.2,7.6){\circle*{1}}
\put(4,10){$\tilde v^\ell$}
\put(17,10.5){$\gamma$}
\put(18.8,7.1){\vector(1,0){1}}
\bezier{300}(10.5,7.5)(12.2,9.7)(13,7.5)
\bezier{300}(13,7.5)(14,6.5)(15,7.5)
\bezier{300}(15,7.5)(16.5,9.7)(17.5,7.5)
\bezier{300}(17.5,7.5)(18.5,6.5)(19.5,7.5)
\bezier{300}(19.5,7.5)(20.5,9.7)(21.5,7.5)
\bezier{300}(21.5,7.5)(22.5,6.5)(23.5,7.5)
\bezier{300}(23.5,7.5)(24.5,9.7)(25.5,7.5)
\put(25.7,7.3){\circle*{1}}
\put(38,14){$k_1$}
\put(25.7,7.3){\line(3,2){11.5}}
\put(25.7,7.3){\vector(3,2){7.5}}
\put(38,-1){$k_2$}
\put(25.7,7.3){\line(3,-2){11.5}} 
\put(25.7,7.3){\vector(3,-2){7.5}}
\end{picture}
\vskip 2cm
\parbox{4.1 in}{\caption{Feynman diagram for a Drell-Yan 
process mediated by vector mesons within our model.}}
\end{center}\label{Feynman}
\end{figure}
\vskip .5cm



\section{Holographic description of dilepton production}

We calculate the contribution to dilepton production represented in the Feynman diagram of Figure 2. 
This corresponds to the interaction of two protons with momenta $P_1$ and $P_2$  through the exchange of vector mesons $v^n$ and $v^m$. These vector mesons combine into another vector meson $v^{\ell}$  that decays  into
a time-like photon that eventually gives rise to a lepton pair. At lowest order, the final state corresponding to each proton is one excited baryon. These final baryons  are not measured, so the hadronic tensor 
$W^{\mu \nu}$ involves the sum over all possible baryonic states $X$ and $Y$. Here we will consider only final states
$X$ and $Y$ with just one spin 1/2 baryon each.

\begin{figure}
\begin{center}
\vskip 3cm
\begin{picture}(0,0)(35,0)
\setlength{\unitlength}{0.035in}
\rm
\thicklines 
\put(-20.3,12.4){\circle*{1}}
\put(-25,0){$\tilde v^n$}
\put(-20,12.2){\line(1,-4){5.3}}
\put(-20.5,12.2){\line(1,-4){5.3}} 
\put(-14.8,-9.2){\circle*{1}}
\put(-25,-20){$\tilde v^m$}
\put(-15,-9.1){\line(1,0){16.5}}
\put(-15,-9.6){\line(1,0){16.5}}
\put(1.7,-9.2){\circle*{1}}
\put(-7,-16){$\tilde v^\ell$}
\put(-19.6,-27.3){\line(1,4){4.5}}
\put(-19.1,-27.3){\line(1,4){4.5}} 
\put(-19.3,-27.35){\circle*{1}}
\put(53.45,12.4){\circle*{1}}
\put(53,0){$\tilde  v^{\bar n}$}
\put(53.2,12.2){\line(-1,-4){5.3}}
\put(53.7,12.2){\line(-1,-4){5.3}} 
\put(31.,-9.){\circle*{1}}
\put(53,-20){$\tilde  v^{\bar m}$}
\put(31.3,-8.9){\line(1,0){16.6}}
\put(31.3,-9.3){\line(1,0){16.9}}
\put(48.1,-9.){\circle*{1}}
\put(36,-16){$\tilde  v^{\bar \ell}$}
\put(53,-27.3){\line(-1,4){4.5}}
\put(52.5,-27.3){\line(-1,4){4.5}} 
\put(52.75,-27.35){\circle*{1}}
\put(10,25){$q$}
\put(6,24){\vector(1,-1){6}}
\put(-11,27){$\mu$}
\multiput(13,5)(5,-5.2){7}{
\bezier{300}(-12,17.2)(-15.4,15.7)(-14.8,19.4)
\bezier{300}(-14.8,19.4)(-13.6,23.8)(-18,22)}
\put(21,25){$q$}
\put(21,18){\vector(1,1){6}}
\put(40,26){$\nu$}
\multiput(-28,-26)(5,5){7}{
\bezier{300}(29.7,16.8)(28.4,21)(32.2,19.4)
\bezier{300}(32.2,19.4)(36.4,17.2)(34.8,21.6)}
\put(-34,16){$P_1$}
\put(-31.2,12.5){\line(1,0){100}}
\put(-36,12.5){\vector(1,0){9}}
\put(-8,12.5){\vector(1,0){3}}
\put(38,6){$P_X$}
\put(38,12.5){\vector(1,0){3}}
\put(-8,6){$P_X$}
\put(59,12.5){\vector(1,0){3}}
\put(65,16){$P_1$}
\put(-34,-34){$P_2$}
\put(-31.2,-27.5){\line(1,0){100}}
\put(-36,-27.5){\vector(1,0){9}}
\put(15,-34){$P_Y$}
\put(15,-27.5){\vector(1,0){3}}
\put(59,-27.5){\vector(1,0){3}}
\put(65,-34){$P_2$}
\end{picture}
\vskip 3cm
\parbox{6.1 in}{\caption{Feynman diagram for the 
proton-proton-photon forward scattering.}}
\end{center}\label{FeynmanCompton}
\end{figure}
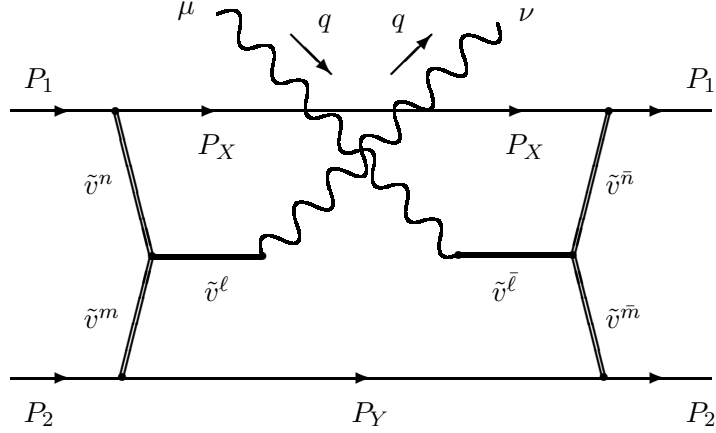
\vskip .5cm



\subsection{Scattering amplitude }

The optical theorem relates the  total cross section of proton-proton-photon scattering to the  forward scattering amplitude 
represented in Figure 3. As a consequence, the structure functions $W_1, W_2, W_3, W_4$ associated with the  
hadronic tensor $W^{\mu \nu}$ in eq. (\ref{Wstructure}) can be obtained from the imaginary part of the forward scattering tensor $T^{\mu \nu}$.

From the diagram of Figure 3 and the Feynman rules that come from the 4-d effective Lagrangians of the previous section we find 
the amplitude
\beqa
W^{\mu \nu} &=& {\rm Im} T^{\mu\nu} = \frac {1}{4(2\pi^2)} f^{abc}f^{abc} \sum_{m, n, \bar m, \bar n} \sum_{n_X,n_Y} 
\int\frac{d^3\vec P_X}{2 \sqrt{{\vec P}_X^2 + M_{n_X}^2}} \int\frac{d^3\vec P_Y}{2 \sqrt{{\vec P}_Y^2 + M_{n_Y}^2}} \cr\cr &\times& \delta^4 \Big( P_1 + P_2 - P_X - P_Y -q \Big)
 \sum_{S_{H1}} \Big( \bar u^1_{(1)} \Gamma_{\bar\alpha}(\bar n ,n_X) [ \gamma_\lambda \, P_X^\lambda + M_{n_X} ] 
\Gamma_{\alpha}(n ,n_X) u^1_{(1)} \Big) \cr
&\times& 
\sum_{S_{H2}} \Big( \bar u^1_{(2)} \Gamma_{\bar\beta}(\bar m ,n_Y) [ \gamma_\sigma \,P_Y^\sigma + M_{n_Y} ] 
\Gamma_{\beta}(m ,n_{Y}) u^1_{(2)} \Big) 
\cr
&\times& \Delta^{\alpha\alpha'}(P_1-P_X , m^2_n) \,\Delta^{\bar\alpha\bar\alpha'}(P_1-P_X , m^2_{\bar n})
\,\Delta^{\beta\beta'}(P_2-P_Y , m^2_m )\,\Delta^{\bar\beta\bar\beta'}(P_2-P_Y , m^2_{\bar m}) \cr \cr
&\times& {\cal C}_{\alpha'\beta'\tau} \, {\cal C}_{\bar\alpha'\bar\beta'\bar\tau}   
\sum_{\ell }  g_{v^nv^mv^{\ell}} g_{v^{\ell}}  
\Delta^{\mu \tau}(q , m_\ell) 
\sum_{ {\bar \ell}}  g_{v^{\bar n}v^{\bar m} v^{\bar \ell}} g_{v^{\bar \ell}} \,
\, \Delta^{\mu \bar \tau}(q , m^2_{\bar \ell})
\label{Amplitude}
\eeqa

\noindent where $u^1_{(1)}$ and $u^1_{(2)}$ are the spinors representing the initial protons with momenta $P_1$ and $P_2$ and 
\beqa 
\Gamma_{\alpha}( n ,n_X) &=& \gamma_\alpha [ g^+_{\bar u^{n_X} v^n u^1} {\cal P}_+ + g^-_{\bar u^{n_X} v^n u^1} 
{\cal P}_- ]\cr\cr
\Delta^{\sigma \rho}( P, m_i^2) &=& \frac{1}{P^2 + m_i^2 } \left [ \eta^{\sigma \rho}   - (1-\xi) \frac{P^\sigma P^\rho }{P^2 + \xi m_i^2 } \right ] \,. \cr\cr
{\cal C}_{\alpha'\beta'\tau} &=& \eta_{\alpha'\beta'} ( P_1 -P_X - P_2 +P_Y)_\tau \cr\cr 
&+& 
\eta_{\beta'\tau } ( P_2 -P_Y + q)_{\alpha'} 
-  \eta_{\alpha'\tau } ( P_1 -P_X + q)_{\beta'}
\eeqa
Note that the denominators of the fermionic propagators do not appear in eq.  (\ref{Amplitude}) since they are transformed
into delta functions  when one takes the imaginary part of the forward scattering tensor. These delta functions impose 
on-shell conditions on the momenta $P_X=(E_X, \vec{P}_X)$ and $P_Y=(E_Y, \vec{P}_Y)$, implying that 
$E_X=\sqrt{\vec{P}_X^2 + M_{n_X}^2}$ and $E_Y=\sqrt{\vec{P}_Y^2 + M_{n_Y}^2}$.

\medskip 

Summing over the spin of the initial hadron 1 we find
\beqa
&&\hskip-.5cm\sum_{S_{H1}}  \bar u^1_{(1)} \Gamma_{\bar\alpha}(\bar n ,n_X) [ \gamma_\lambda p_X^\lambda + M_{n_X} ] 
\Gamma_{\alpha}(n ,n_X) u^1_{(1)} \cr
&=&  P_1^\lambda P_X^\sigma \Big[ g^+_{\bar u^{n_X} v^{\bar n} u^1}   
g^+_{\bar u^{n_X} v^n u^1} \tr \Big( \gamma_\lambda \gamma_{\bar\alpha} \gamma_{\sigma} \gamma_{\alpha}
{\cal P}_+ \Big) + g^-_{\bar u^{n_X} v^{\bar n} u^1}   
g^-_{\bar u^{n_X} v^n u^1} \tr \Big( \gamma_\lambda \gamma_{\bar\alpha} \gamma_{\sigma} \gamma_{\alpha}
{\cal P}_- \Big) \Big] \cr\cr &+& \frac12 M_1 M_{n_X} \Big[ g^+_{\bar u^{n_X} v^{\bar n} u^1} g^-_{\bar u^{n_X} v^n u^1} 
+ g^-_{\bar u^{n_X} v^{\bar n} u^1} g^+_{\bar u^{n_X} v^n u^1} \Big] \tr \Big(\gamma_{\bar\alpha} \gamma_{\alpha}  \Big) 
\eeqa

\noindent where the traces over gamma matrices are 
\beqa
\tr \Big( \gamma_\lambda \gamma_{\bar\alpha} \gamma_{\sigma} \gamma_{\alpha} {\cal P}_\pm \Big)&=& 2 [ \eta_{\lambda{\bar\alpha}} \eta_{\sigma\alpha}
 - \eta_{\lambda \sigma } \eta_{{\bar\alpha} \alpha} + \eta_{\lambda \alpha} \eta_{{\bar \alpha} \sigma}  ] \, + \dots \cr
\tr \Big(  \gamma_{\bar\alpha} \gamma_{\alpha}   \Big) &=& - 4 \eta_{\bar\alpha \alpha} 
\eeqa
\noindent and the dots represent the imaginary term  $\mp 2 i \epsilon_{ \lambda{\bar \alpha} \sigma\alpha}$ whose contribution is negligible. For the sum over the spin of hadron 2 we have a similar result. 

In order to perform the integration in the three momenta ${\vec P}_X$ and $ {\vec P}_Y$ we choose a frame where 
${\vec P}_1 +  {\vec P}_2 - \vec q = 0$. That means, we work on the center of momentum frame of final hadrons.
However, it should be stressed that the results for the structure functions $W_1,W_2,W_3, W_4 $ are frame independent. 
Then the integral takes the form
\beqa  
&&\int\frac{d^3\vec P_X}{2 E_X } \int\frac{d^3\vec P_Y}{2 E_Y }\,\delta^3 \Big( {\vec P}_X + {\vec P}_Y \Big) \delta \Big( E_X + E_Y - E_F \Big) \times  ... \cr\cr
&&\, =\,  \int\frac{d^3\vec P_X}{4 E_X \sqrt{ E_X^2 - M_{n_X}^2 + M_{n_Y}^2  }} \delta \Big( E_X + \sqrt{ E_X^2 - M_{n_X}^2 + M_{n_Y}^2  } - E_F \Big) \times  ... \qquad 
\eeqa

\noindent  where $E_F \equiv E_1 + E_2 - q_0 $ is the energy available for the final hadrons. We can express the three momentum volume in terms of the energy as
\beq
{d^3\vec P_X} = E_X \sqrt{ E_X^2 - M_{n_X}^2 } \sin \theta_X dE_X d\theta_X d\phi_X \,.
\eeq

\noindent Here $\theta_X $ and $\phi_X $ are spherical angular coordinates for the vector $\vec P_X$.
This way, the delta function can be integrated leading to the condition 
\beq
E_X = \frac{1}{2 E_F} \Big( E_ F^2 +  M_{n_X}^2 - M_{n_Y}^2 \Big)\,.
\eeq

\noindent Then the momentum integrals in eq. (\ref{Amplitude}) become just  angular integrations, i.e. 
\beq
\frac14 \int_0^\pi d\theta_X \int_0^{2\pi} d\phi_X  \sin \theta_X \, \frac{\sqrt{ E_X^2 - M_{n_X}^2 } }{E_F} \left (  ... \right ) \,\,.
\eeq

\noindent The integrand depends on the internal momenta $P_X$ and $ P_Y $ (in the Feynman graph of Figure 3 ). 
In the center of momentum frame of final hadrons, these momenta are expressed as:
\beqa   
P_X &=& (E_X ,  \vert \vec{P}_X \vert\sin\theta_X \cos\phi_X , 
\vert \vec{P}_X \vert \sin\theta_X \sin\phi_X , \vert \vec{P}_X \vert \cos\theta_X )
\cr
P_Y &=& (E_F - E_X , - {\vec P}_X )  
 \,.
\eeqa
where $\vert \vec{P}_X \vert = \sqrt{ E_X^2 - M_{n_X}^2  }$. 
The integrals in $\theta_X $ and $\phi_X$ are performed for each term in the sum over vector mesons and fermions. 

The amplitude defined in eq. (\ref{Amplitude}) involves sums over all intermediate vector mesons and final baryons. 
For the final baryons  $X$ and $Y$, there is a physical condition that sets a natural cut off for the sums, 
since they are on shell particles. The available energy for them is 
$E_F \equiv E_1 + E_2 - q_0 $, so, the final states must satisfy:
\beq 
M_{n_X} +  M_{n_Y} \le E_F\,.
\eeq

For the vector mesons, there is no physical cut off since they are off-shell. In our numerical approach, we used the convergence of the sums as the criterion to define how many vector mesons should be summed in the internal lines.  We computed the sums over the indices $m, \, n,\, {\bar m}, {\bar n} $ from 1 to some integer value $n_{max} $ starting with $n_{max} = 8 $ and increasing its value.  We did the same for the indices $\ell,\, {\bar \ell} $ taking the corresponding sums from 1 to some integer $n_{max2}$. We found a good convergence at $n_{max} = 15 $ and $n_{max2} = 25$.
More precisely, adding one more term to each vector meson sum, that means taking $n_{max} = 16 $ and 
$n_{max2} = 26$ we found a relative variation smaller than $10^{-6}$ in expression (\ref{Amplitude}).
So, we used $n_{max} = 15 $ and $n_{max2} = 25$ in all our calculations. The relative error in the angular parameters, associated with this cut off,  is of order $10^{-6}$.  
  
In our amplitude (\ref{Amplitude}) we wrote for the propagators of all vector mesons of momentum $P$ and mass $m_i$  the expression
\beq  
\Delta^{\mu\nu}(P , m^2_i) \,=\, \frac{1}{P^2 + m_i^2 } \left [ \eta^{\mu \nu}   - (1-\xi) \frac{P^\mu P^\nu }{P^2 + \xi m_i^2 } \right ] \,.
\eeq
This propagator depends on the 5-d gauge fixing parameter $\xi$ so depending on the choice of $\xi$ we may have different effective 4-d models.  There are three common gauge choices : i)  $\xi \to \infty$ which leads to transverse currents but is hard to implement  
at high energies, ii)  $\xi = 1$ which does not lead to transverse currents unless we turn on Goldstone bosons  
and iii) $\xi=0$ which leads to transverse currents and has a nice high energy behavior.

 As we explained above, the calculation of the dilepton cross section involves momentum integration and several sums over fermion and vector meson masses. For this reason, we found convenient to work with the gauge $\xi=0$ where the propagator leads to transversality of the hadronic tensor and has a good numerical behavior in the sense that it reduces to a massless propagator in the limit of large  $P^2$. We also made some numerical tests with the gauge $\xi \to \infty$ but in that case the mass dependence on the numerator implies a huge running time making very hard to guarantee convergence of the different sums.

\subsection{Numerical set up and results}

\begin{table}
\begin{tabular}{|cccccc|ccc|cc|}
\hline 
\multicolumn{6}{|c|}{ {\rm Kinematical \, regimes}} & \multicolumn{3}{c|}{ \rm Our \,results} & \multicolumn{2}{c|}{ \rm Perturb. \,results }\\
\hline  
$p$ & $q_2$ & $q_3$ & $\sqrt{s}$ & $q_T$ & $x_F$ & $\lambda$ & $\mu$ & $\nu$ & $\lambda_{pert}$ & $\nu_{pert}$ \\  \hline\hline
\, 25.8  \,&\, 1.61  \,&\, 11.3  \,&\,  38.8 \,&\, 1.47  \,&\, 0.373 \,&\, 0.76 \,&\,  0.069 \,&\, -0.084  \,&\, 0.960  \,&\, 0.0201 \,  \\
\, 25.6 \,&\, 1.29  \,&\, 10.8  \,&\,  38.9 \,&\, 1.20  \,&\, 0.361 \,&\, 0.76 \,&\,   0.066 \,&\, -0.074  \,&\, 0.973  \,&\, 0.0136 \,  \\
\, 24.8 \,&\, 0.937 \,&\,  9.69 \,&\,  38.9 \,&\, 0.900 \,&\, 0.331 \,&\, 0.73 \,&\,   0.032 \,&\,  0.069 \,&\, 0.985  \,&\, 0.00770 \, \\
\, 24.1 \,&\, 0.565 \,&\,  8.40 \,&\,  38.9 \,&\, 0.601 \,&\, 0.294 \,&\, 0.70 \,&\,   0.034 \,&\, -0.0024 \,&\, 0.993  \,&\,  0.00346 \, \\
\, 22.7 \,&\, 0.162 \,&\,  6.02  \,&\, 38.9 \,&\, 0.302 \,&\, 0.218 \,&\, 0.57 \,&\,   0.019 \,&\,  0.030 \,&\, 0.998  \,&\, 0.000876 \, \\ \hline
\end{tabular}
\caption{ Angular parameters calculated from our model and from the perturbative approach, for $m_\gamma^2 = 104 \, {\rm GeV}^2 $. The kinematical variables $p, q_2, q_3, \sqrt{s}, q_T$ are expressed in GeV. Note that $\mu_{pert}$ is not known, since it depends on the unkown parton distribution functions \cite{Boer:2006eq}.
\label{vectortableconstants1}}
\end{table}

\begin{table}
\begin{tabular}{|cccccc|ccc|cc|}
\hline 
\multicolumn{6}{|c}{ {\rm Kinematical \, regimes}} 
& \multicolumn{3}{|c|}{ \rm Our  \,results} 
& \multicolumn{2}{c|}{ \rm Perturb. \,results } \\
 \hline 
$p$ & $q_2$ & $q_3$ & $\sqrt{s}$ & $q_T$ & $x_F$ & $\lambda$ & $\mu$ & $\nu$ & $\lambda_{pert}$ & $\nu_{pert}$ \\ \hline\hline
 \, 25.8 \,&\,  1.53   \,&\, 11.3  \,&\, 38.9 \,&\,  1.50  \,&\, 0.398 \,&\, 0.60 \,&\,  0.11  \,&\, -0.15   \,&\, 0.925 \,&\, 0.0377 \, \\
 \, 25.5 \,&\,  1.20   \,&\, 10.8  \,&\, 38.9 \,&\,  1.20  \,&\, 0.386 \,&\, 0.61 \,&\,  0.085  \,&\, -0.11  \,&\, 0.951 \,&\, 0.0246 \, \\
 \, 24.8 \,&\,  0.859  \,&\, 9.69  \,&\, 38.9 \,&\,  0.900 \,&\, 0.355 \,&\, 0.57 \,&\,  0.070  \,&\, -0.038  \,&\, 0.972 \,&\, 0.0141 \, \\
 \, 24.1 \,&\,  0.501  \,&\, 8.40  \,&\, 38.9 \,&\,  0.600 \,&\, 0.316 \,&\, 0.49 \,&\,  0.057  \,&\,  0.025 \,&\, 0.987 \,&\, 0.00633 \, \\
 \, 22.6 \,&\,  0.0808 \,&\, 6.02  \,&\, 38.9 \,&\,  0.302 \,&\, 0.236 \,&\, 0.32 \,&\,  0.021  \,&\,  -0.013  \,&\, 0.997      \,&\,    0.00162 \,   \\  \hline
\end{tabular}
\caption{ Angular parameters calculated from our model and from the perturbative approach, for $m_\gamma^2 = 56.3 \, {\rm GeV}^2 $. The kinematical variables $p, q_2, q_3, \sqrt{s}, q_T$ are expressed in GeV. Note that $\mu_{pert}$ is not known, since it depends on the unknown parton distribution functions \cite{Boer:2006eq}.
\label{vectortableconstants2}}
\end{table}

\begin{figure}
\includegraphics[bb=0 0 300 216,width=6.99cm,height=5.04cm,keepaspectratio]{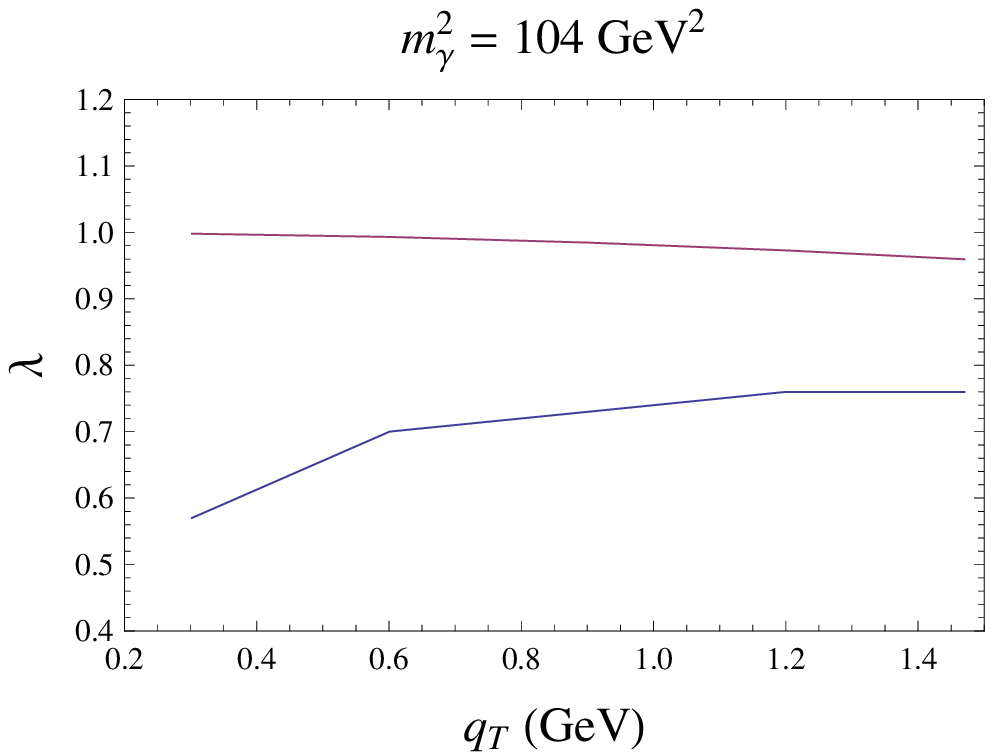}
\includegraphics[bb=0 0 300 213,width=7.11cm,height=5.04cm,keepaspectratio]{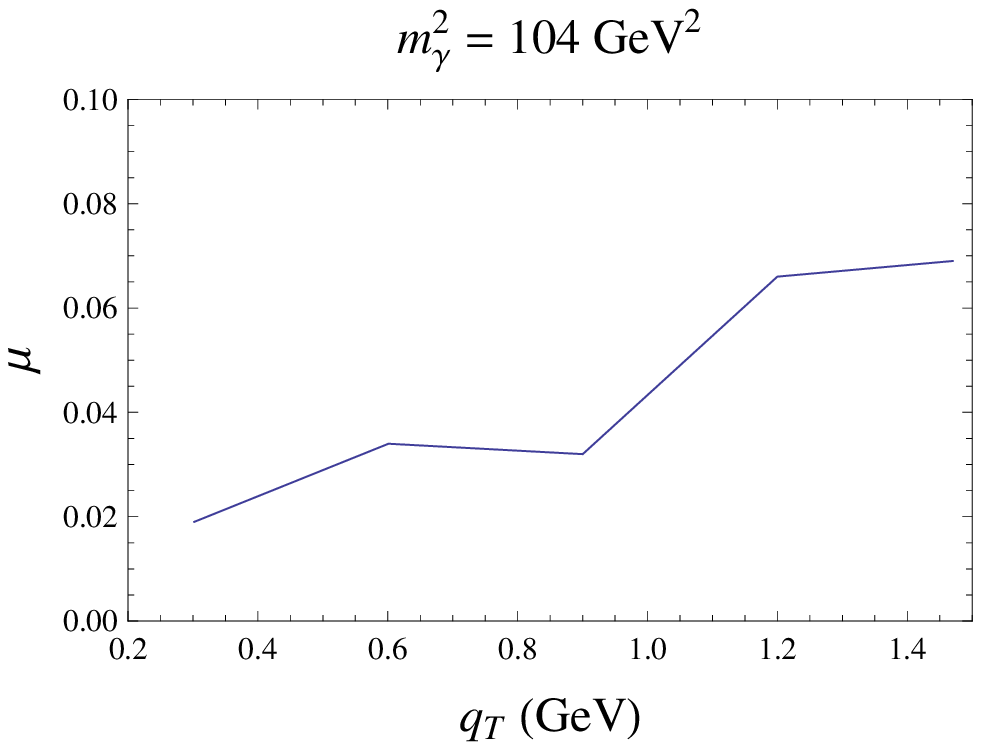}
\includegraphics[bb=0 0 300 209,width=7cm,height=4.88cm,keepaspectratio]{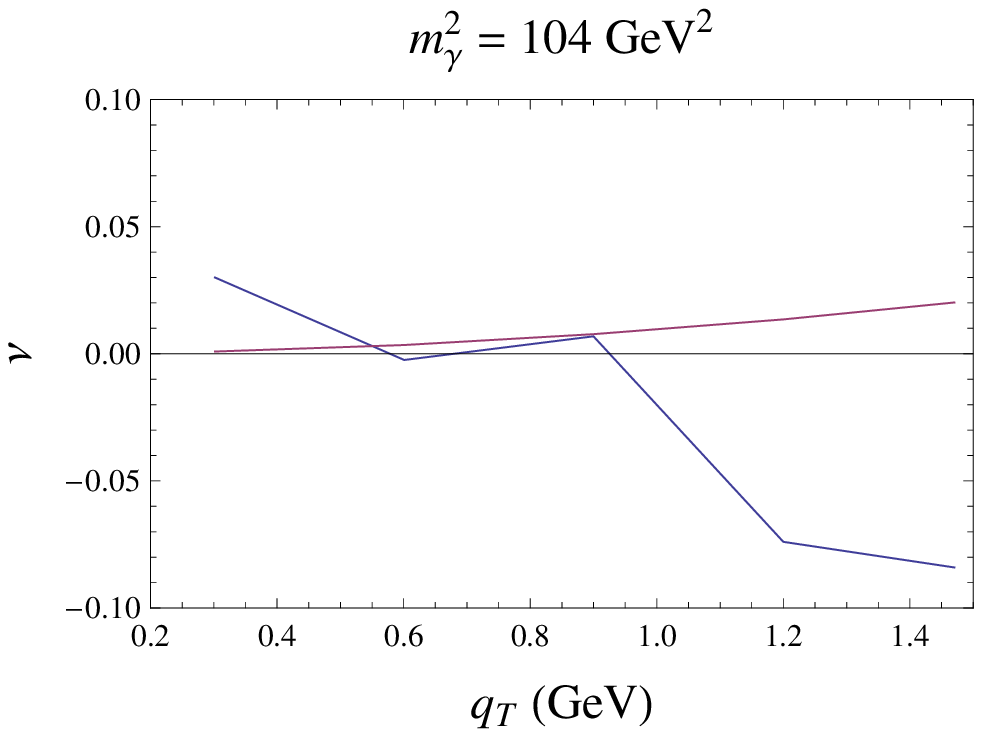}
\caption{\label{graficos} We plot our results (blue lines) and the perturbative predictions (red lines) for the angular parameters $\lambda,\mu,\nu$ for $m_\gamma^2=104{\rm GeV}^2 $.  Note that $\mu$ is not compared with the corresponding perturbative result $\mu_{pert}$ since the later it is not known, thanks to its depence on the unknown parton distribution functions \cite{Boer:2006eq}.} 
\end{figure}

\begin{figure}
\includegraphics[bb=0 0 300 216,width=7cm,height=5.05cm,keepaspectratio]{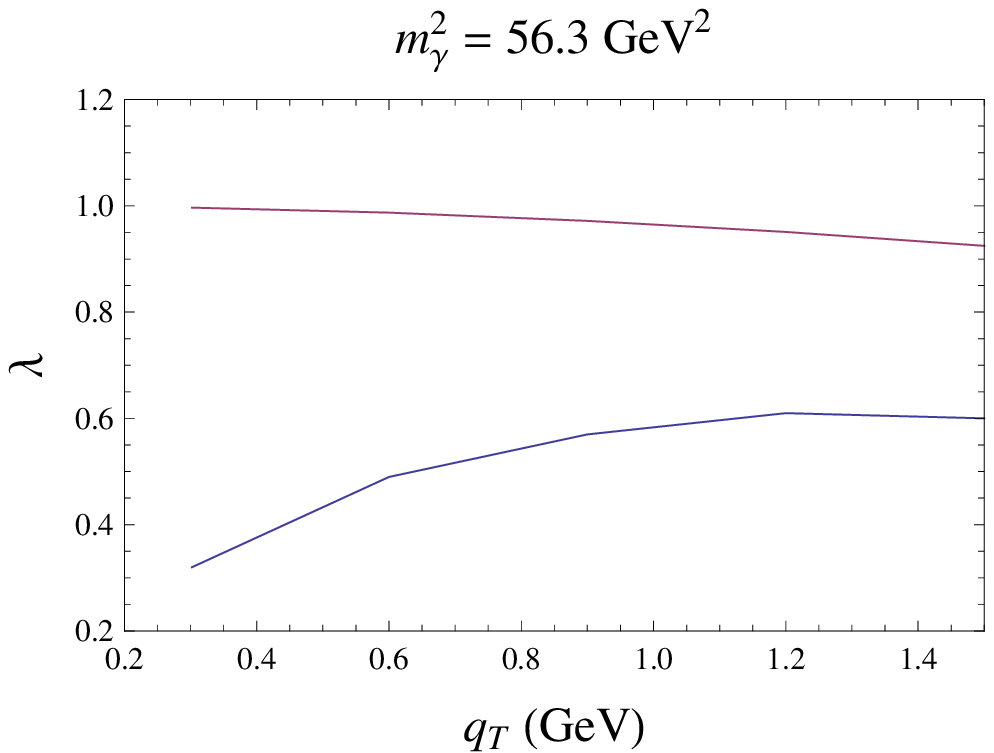}
\includegraphics[bb=0 0 300 213,width=7cm,height=4.96cm,keepaspectratio]{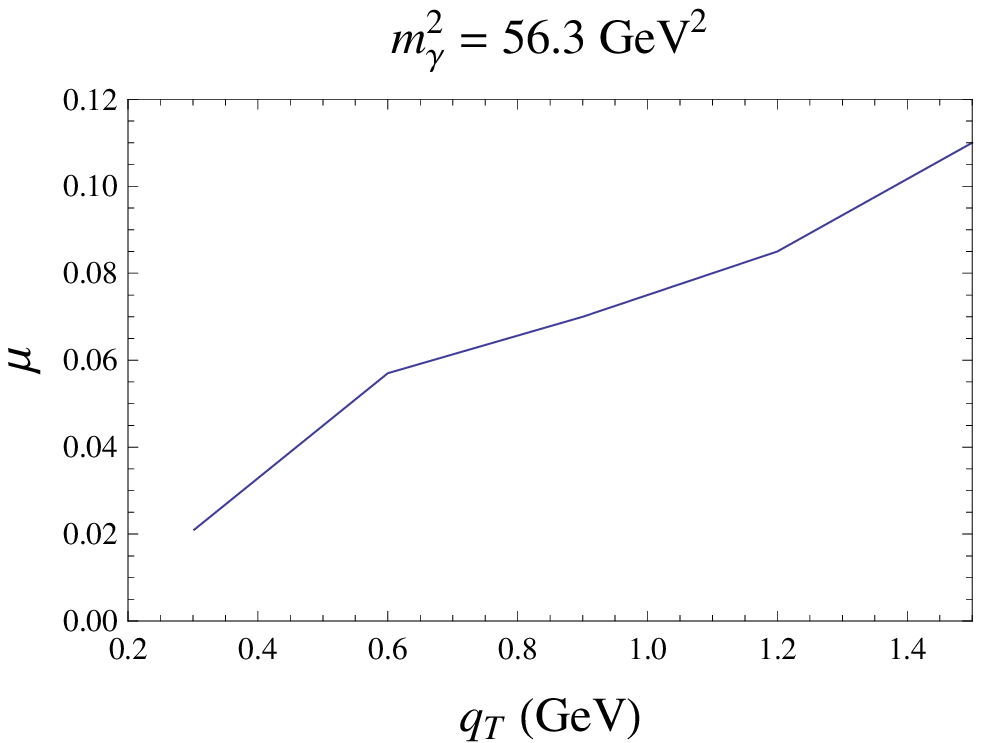}
\includegraphics[bb=0 0 300 212,width=7cm,height=4.95cm,keepaspectratio]{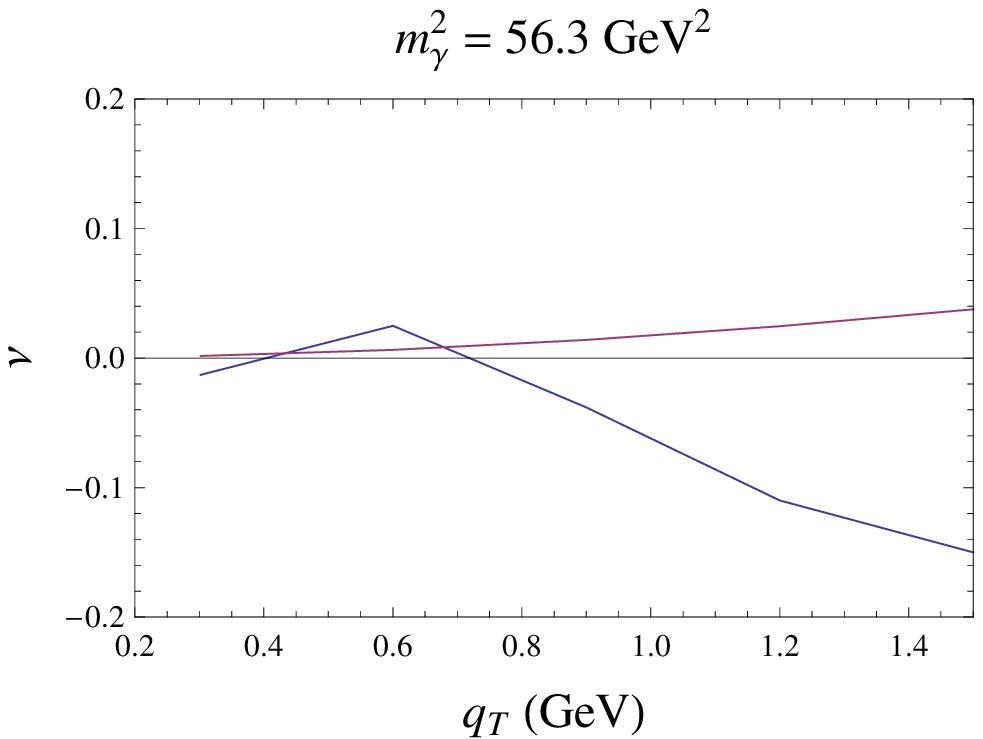}
\caption{\label{graficos2} We plot our results (blue lines) and the perturbative predictions (red lines) for the angular parameters $\lambda,\mu,\nu$ for $m_\gamma^2 = 56.3 \, {\rm GeV}^2 $.  Note that $\mu$ is not compared with the corresponding perturbative result $\mu_{pert}$ since the later it is not known, thanks to its dependence on the unknown parton distribution functions \cite{Boer:2006eq}.}
\end{figure}

We fix the size of the hard wall AdS slice $z_0 = 1/\Lambda $,  by fitting the mass of the $\rho$ meson $m_{\rho} = 0.776 $ GeV. This gives $ \Lambda = 0.323 $ GeV . 
Concerning the parameter $\kappa R $ that, as discussed in section III, shows up in all the couplings, one can see from the form of the hadronic tensor eq. 
(\ref{Amplitude}) that it will contribute just to a common multiplicative factor. So, the same factor will appear in all the structure functions. In particular, the helicity structure functions defined in eq. (\ref{Whelicity}) will share the same multiplicative factor of $\kappa R $. Thus, the parameters $\lambda,\mu,\nu$, defined as ratios of linear combinations of these objects, will be independent of $\kappa R $ and we do not need to fix it.

In our frame, of the center of mass of the final hadrons,  the protons and the photon momenta can be written as
\beqa   
P_1 &=& (\sqrt{ p^2 + M_1^2 }\,, 0\,,0,\,p) \quad ; \quad q = ( \sqrt{ m_\gamma^2 + q_2^2 + q_3^2}\,, 0, q_2, 
\,q_3) \cr
P_2 &=& (\sqrt{ q_2^2 + (q_3 - p)^2 + M_1^2 }\,, 0\,, q_2 ,q_3 - p\,)\,, 
\eeqa

\noindent where $M_1 = 3.83 \Lambda = 1.2$ GeV  is the ``proton'' mass in the model, once the infrared scale was fixed by the mass of the $\rho$ meson. We analyzed different kinematical regimes,  contained in the region analyzed from experimental data in ref. \cite{Zhu:2008sj}. Each one was defined by a choice of the values of $ p,\, q_2,\, q_3 $ and $ \, m_\gamma $. From these quantities we obtained the corresponding values for $ \,\sqrt{s}\,, \,  q_T $ and 
$ x_F \,\equiv 2\frac{q_p}{\sqrt{s}}\, $ where $q_T$ and $q_p$ were defined in eq. (\ref{qes}). 

The numerical calculation where performed using the package ``Mathematica''. 
From eq. (\ref{Amplitude}) we computed the diagonal elements 
of the hadronic tensor $W^{\mu \nu}$. Using these results it is possible to invert 
eq. (\ref{Wstructure}) and obtain the structure functions  $W_1, W_2, W_3, W_4$. 
Then from  eq.  (\ref{LamTungeq}) we obtain the helicity structure functions in the 
Collins-Soper frame from which we calculate the angular parameters $\lambda, \mu , \nu$ defined in eq. (\ref{lambdamunu}).

 We chose kinematical regimes in the region analyzed in ref.\cite{Zhu:2008sj}.
They covered the range $ 20 < m_{\mu\mu}^2 < 225 \,{\rm GeV}^2 $, so we decided to test, as representatives of this region, the values $ m_\gamma^2 \equiv m_{\mu\mu}^2 = 56.2 \, {\rm GeV}^2$ and  $ m_\gamma^2 =  104 \,{\rm GeV}^2$.
For the transverse momentum $q_T$ they covered the region  $0 < q_T < 4 \,{\rm GeV}$, so we chose values in  the non perturbative range $ 0.3 < q_T <  1.5 \,{\rm GeV}$. 

In Tables \ref{vectortableconstants1} and \ref{vectortableconstants2}, we show our choices of kinematical regimes and the corresponding results for the angular parameters from our model and from the perturbative expressions 
for $\lambda$ and $\nu$ (\ref{perturb_results}), since the perturbative expression for $\mu$ depends on the parton distribution functions which are not known. 
In Table \ref{vectortableconstants1}  we fixed $m_\gamma^2 = 1000 \Lambda^2 \,=\, 104 \, {\rm GeV}^2 $ while in Table \ref{vectortableconstants2},  $m_\gamma^2 = 539 \Lambda^2 \,=\, 56.2 \, {\rm GeV}^2 $.

We also present plots comparing our results  for $\lambda$, $\mu$, and $\nu$ with the perturbative ones $\lambda_{pert}$ and $\nu_{pert}$ in Figures {\bf 4} and {\bf 5}. We see that our model predicts a decrease in the value of $\lambda$ as $q_T$ decreases. This contrasts with the perturbative expectations of $\lambda\to 1$ when $q_T\to 0$. The mean value  $<\lambda> =$ 0.85 obtained from the available experimental data is lower than the perturbative and greater than our model results. 

Now we estimate the error associated with the choice of boundary conditions for the fermions by calculating 
the angular parameters using the alternative boundary condition $\tilde \phi^n (z_0)= 0$. For this purpose, we 
calculated again all the fermionic masses and coupling constants for this alternative condition. 
We compare in table III  the angular parameters obtained using the original boundary condition with those
obtained using the alternative boundary conditions. From this table we can have some estimative of the 
relative errors associated with the choice of boundary conditions. We find
\begin{equation}
\frac{\delta\lambda}{\lambda} \sim 0.04 \,-\, 0.06 \,, \qquad\qquad \frac{\delta\mu}{\mu} \sim 0.6 \, -\, 0.8 \,, \qquad\qquad \frac{\delta\nu}{\nu} \sim  0.06 \, - \, 1.4  \,.  \qquad 
\end{equation}
These results show that the parameter $\lambda$ has a low sensitivity to the choice of boundary conditions for the fermions. This is the largest angular parameter and the one for which the predictions of our model are robust and closer to both experimental and perturbative results.  
The experimental results for $\mu$ and $\nu$ are much smaller and present oscillations.
In our model these parameters are very sensitive to the choice of fermionic boundary conditions.
However, the absolute values of our parameters are of the same order of the experimental ones.  
 
\begin{table}
\begin{tabular}{|cccccc|ccc|ccc|}
\hline 
\multicolumn{6}{|c|}{ {\rm Kinematical \, regimes}} & \multicolumn{3}{c|}{ \rm Original B.C.} & \multicolumn{3}{c|}{ \rm Alternative B.C. }\\
\hline  
$p$ & $q_2$ & $q_3$ & $ m_\gamma^2 $ & $q_T$ & $ \sqrt{s} $ & $\lambda$ & $\mu$ & $\nu$ & $\lambda_{a}$ & $\mu_{a}$ & $\nu_{a}$ \\  \hline\hline
\, 25.8  \,&\, 1.61  \,&\, 11.3  \,&\,  104 \,&\, 1.47  \,&\, 38.8 \,&\, 0.76 \,&\,  0.069 \,&\, -0.084  \,&\, 0.73  \,&\, 0.13 \,& \, -0.079  \\
 \, 22.6 \,&\,  0.0808 \,&\, 6.02  \,&\, 56.3 \,&\,  0.302 \,&\, 38.9 \,&\, 0.32 \,&\,  0.021  \,&\, -0.013  \,&\, 0.34     \,&\,    0.048  \, & \, -0.072 \\ \hline
\end{tabular}
\caption{ Angular parameters calculated with the original boundary condition $\phi^n (z_0) = 0$ and with 
the alternative boundary condition   ${\tilde \phi}^n (z_0) = 0$. The kinematical variables $p, q_2, q_3, \sqrt{s}, q_T, m_\gamma $ are expressed in GeV. \\
\label{vectortableconstants3}}
\end{table}

\section{Conclusions}

As we mentioned in the introduction, the perturbative QCD approach provides a good 
approximation for the dilepton production cross section when the dilepton transverse momentum $q_T$ is large compared to the virtual photon mass $m_\gamma$. In this approach the hadronic tensor $W^{\mu \nu}$ depends, through collinear convolution, on the generalized parton distribution functions and partonic cross sections. The dominant processes  are the Drell Yan quark annihilation ($ q \bar q \to \gamma^\ast$), the quark annihilation with the emission of a gluon ($q \bar q \to \gamma^\ast g$) 
and the quark gluon scattering $q g \to \gamma^\ast q$ (where $\gamma^\ast$ is the virtual photon). However, in the limit of small dilepton transverse momentum ($q_T \ll m_\gamma$) the perturbative approach suffers from large logarithmic corrections \cite{Boer:2006eq}.
Since the data extracted from recent experiments (like the FNAL E866/NuSea Collaboration) include this problematic regime in which the effective coupling is large it is worth to explore alternative approaches to the regime of small $q_T$.

In this article we presented a holographic model where dileptons are produced in proton-proton collisions by the exchange of vector mesons. Our approach deals only with dileptons that are produced from time-like virtual photons.
The model describes only processes where, after the collision, each proton transforms into a single baryon of spin 1/2. It would be interesting to extend this study in order to include the production of more general final hadronic states, like baryons of spin 3/2. 

We applied our model to estimate the cross section parameters  $\lambda,\mu,\nu$ that characterize the angular distribution of dileptons produced in a proton-proton collision for small $q_T$. 
We compared our results  for $\lambda$ and $\nu$ with those from perturbative QCD. We found that in the region analysed $\lambda$ is smaller than $\lambda_{pert}$ and 
increases with $q_T$ in contrast with the perturbative result. For $q_T\to 0$ the perturbative result is $\lambda_{pert}\to 1$ while our model gives decreasing values for $\lambda$.

As a final remark, it is important to mention that there are other processes that can generate dileptons in a proton-proton collision and may contribute to the  parameters $\lambda,\mu,\nu$. Then it would be interesting to explore possible extensions of the present model to include these other processes.

\bigskip

\noindent {\bf Acknowledgements:} We thank Lingyan Zhu for explaining details of the kinematical regime investigated in ref. \cite{Zhu:2008sj}. We also thank Marcus Torres and Miguel Quartin for helping us with the numerical computations. C.A.B.B would like to acknowledge  the support of the  STFC Rolling Grant ST/G000433/1. The authors are partially supported by CAPES, CNPq and FAPERJ.

\section{Appendix: tables of masses and couplings}

Here we list some masses and couplings for vector mesons and spin 1/2 baryons in the hard wall model.

\begin{table}
\scriptsize
\begin{tabular}{|c|cccccccc|}
\hline
\,$n$\, & $\frac{m_n }{\Lambda}$ &  $\frac{g_{v^n}}{\sqrt{\kappa R} \, \Lambda^2}$ 
& ${\ss\sqrt{\ss \kappa R}} \, g_{v^1v^1v^n}$ & ${\ss\sqrt{\ss \kappa R}} \, g_{v^1v^2v^n}$ & ${\ss\sqrt{\ss \kappa R}} \, g_{v^1v^3v^n}$ & ${\ss\sqrt{\ss \kappa R}} \, g_{v^2v^2v^n}$ & ${\ss\sqrt{\ss \kappa R}} \, g_{v^2v^3v^n}$ & ${\ss\sqrt{\ss \kappa R}} \, g_{v^3v^3 v^n}$  \\ 
 [0.5ex] \hline\hline 
\,1\, &\, 2.4048 &6.5510  &1.0923  &-0.31784 &0.0035072  &0.74819 &-0.34179 &0.71437 \\
\,2\,&\, 5.5201 &22.943  &-0.31784  &0.74819 &-0.34179  &0.12962 &0.40807 & 0.10405\\
\,3\,&\, 8.6537 &45.084  & 0.0035072 &-0.34179 &0.71437  &0.40807 &0.10405 &0.13996 \\
\,4\,&\, 11.791  &71.736  &-0.00039171  &0.0053568 &-0.34723  &-0.37659 &0.36944 & 0.076647\\
\,5\,&\, 14.931 &102.23  &0.000083373  &-0.00074121 &0.0060115 &0.0087557 &-0.38591 & 0.32767\\
\,6\,&\, 18.071 &136.13  &-0.000024843  &0.00018336 &-0.00090370 &-0.0015000 &0.010128 & -0.39707\\
\,7\,&\, 21.212  &173.13  &9.1480\textsf{x}$10^{-6}$ &-0.000061090 &0.00023961 &0.00042793 &-0.0018924 & 0.011905\\
\,8\,&\, 24.352  &212.98  &-3.9015\textsf{x}$10^{-6}$  &0.000024521 &-0.000084592  &-0.00015811 &0.00057805 & -0.0024352\\
\,9\,&\, 27.493  &255.50  &1.8545\textsf{x}$10^{-6}$ &-0.000011201 &0.000035649  &0.000068680 &-0.00022572 & 0.00079693 \\ 
\,10\,&\, 30.635 &300.51  &-9.5825\textsf{x}$10^{-7}$ &5.6307\textsf{x}$10^{-6}$  &-0.000016972 &-0.000033399 &0.00010265 & -0.00032869 \\
\,11\,&\, 33.776 &347.90  &5.2915\textsf{x}$10^{-7}$ &-3.0479\textsf{x}$10^{-6}$ &8.8399\textsf{x}$10^{-6}$  &0.000017666 &-0.000051882 & 0.00015630\\
\,12\,&\, 36.917  &397.55 &-3.0844\textsf{x}$10^{-7}$&1.7504\textsf{x}$10^{-6}$ &-4.9345\textsf{x}$10^{-6}$ &-9.9764\textsf{x}$10^{-6}$ &0.000028362 & -0.000081999\\
\,13\,&\, 40.058  &449.36  &1.8807\textsf{x}$10^{-7}$ &-1.0552\textsf{x}$10^{-6}$ &  2.9110\textsf{x}$10^{-6}$ &5.9383\textsf{x}$10^{-6}$ &-0.000016480 & 0.000046268\\
\,14\,&\, 43.200  &503.25  &-1.1911\textsf{x}$10^{-7}$  &6.6235\textsf{x}$10^{-7}$ &-1.7968\textsf{x}$10^{-6}$ &-3.6912\textsf{x}$10^{-6}$ &0.000010057 & -0.000027626\\
\,15\,&\, 46.341  &559.13 &7.7939\textsf{x}$10^{-8}$ &-4.3029\textsf{x}$10^{-7}$ &1.1518\textsf{x}$10^{-6}$  & 2.3795\textsf{x}$10^{-6}$ &-6.3897\textsf{x}$10^{-6}$ & 0.000017263\\
\hline
\end{tabular}
\caption{Some masses and coupling constants of vector mesons in the hard wall model.
\label{tableconstantsvectormesons}}
\end{table}

\begin{table}
\scriptsize
\begin{tabular}{|c|ccccccc|}
\hline
 $n$ & $\frac{M_n }{\Lambda}$ &  ${\ss\sqrt{\kappa R}}\, g^+_{\bar u^1 v^n u^1}$ &
${\ss\sqrt{\ss \kappa R}}\, g^+_{\bar u^1 v^n u^2}$ & ${\ss\sqrt{\ss \kappa R}}\, g^+_{\bar u^1 v^n u^3}$ & ${\ss\sqrt{\ss \kappa R}}\, g^-_{\bar u^1 v^n u^1}$ & 
${\ss\sqrt{\ss \kappa R}}\, g^-_{\bar u^1 v^n u^2}$ & ${\ss \sqrt{\ss \kappa R}}\, g^-_{\bar u^1 v^n u^3}$  \\ 
\hline\hline
 1&\,5.1356   & 0.92525      & -0.28634    & 0.0063502    & 1.1995      & -0.21837     & -0.0010294     \\
 2&\,8.4172   & 0.22261      & 0.55996     & -0.29692     & -0.55390    & 0.68808      & -0.18644       \\
 3& \,11.620  & -0.49828     & 0.11371     & 0.52173      & 0.056561    & -0.62284     & 0.59639        \\
 4&\, 14.796  & 0.17584      & -0.49341    & 0.078525     & 0.087530    & 0.19693      & -0.63706       \\
 5&\, 17.960  & -0.017595    & 0.20755     & -0.49104     & -0.070083   & 0.040080     & 0.25764        \\
 6&\, 21.117  & 0.0053409    & -0.019083   & 0.22278      & 0.043309    & -0.058728    & 0.014012       \\
 7&\, 24.270  & -0.0022875   & 0.0059035   & -0.019476    & -0.029528   & 0.036904     & -0.052222      \\
 8&\, 27.421  & 0.0011692    & -0.0026014  & 0.0060597    & 0.021530    & -0.025401    & 0.033015       \\
 9&\, 30.569  & -0.00066745  & 0.0013668   & -0.0027067   & -0.016447   & 0.018679     & -0.022778      \\ 
 10&\,33.717  & 0.00041141   & -0.00079982 & 0.0014436    & 0.013002    & -0.014384    & 0.016797       \\
 11&\,36.863  & -0.00026846  & 0.00050392  & -0.00085719  & -0.010551   & 0.011455     & -0.012977      \\
 12&\,40.008  & 0.00018311   & -0.00033522 & 0.00054749   & 0.0087422   & -0.0093591   & 0.010371       \\
 13&\,43.154  & -0.00012940  & 0.00023256  & -0.00036882  & -0.0073665  & 0.0078027    & -0.0085037     \\
 14&\,46.298  & 0.000094153  & -0.00016685 & 0.00025883   & 0.0062949   & -0.0066124   & 0.0071147      \\
 15&\,49.442  & -0.000070205 & 0.00012304  & -0.00018766  & -0.0054432  & 0.0056801    & -0.0060500     \\
\hline
\end{tabular}
\caption{Some masses and coupling constants of fermions in the hard wall model with the boundary condition $\phi^n(z_0)=0$. 
\label{tableconstantsfermions}}
\end{table}

Vector meson masses are obtained from the zeros of Bessel function $J_0(w)$. 
The $g_{v^n}$ coupling constants, calculated using eq. (\ref{gvn}), 
 represent the decay of a vector $v^n$ into a photon.
 The vector meson triple couplings $g_{v^1v^iv^n}$ are calculated from eq. (\ref{gvvv}). 
Some of these results are shown in Table \ref{tableconstantsvectormesons}.

The masses for spin 1/2 baryons are calculated from the zeros of the Bessel functions $J_1(w)$. 
The triple couplings among two baryons and one vector meson are calculated from eq. (\ref{ghhv}).
Some of these results are shown in Table \ref{tableconstantsfermions}.

\end{document}